# Chiral string theories as an interpolation between strings and particles

Matheus Loss Lize

*Instituto de Física Teórica*
*Universidade Estadual Paulista*

Bochen Lyu

*Dept. of Physics and Astronomy*
*Stony Brook University*

Warren Siegel

*CNYITP*
*Stony Brook University*

Yu-Ping Wang

*Dept. of Physics and Astronomy*
*Stony Brook University*

September 27, 2021

### Abstract

A new set of boundary conditions for string propagators is proposed in this paper. The boundary conditions are parametrized by a complex number $\lambda$. Under these new boundary conditions, the left-moving and right-moving modes are treated unequally. Thus, we called them chiral string theories. If $\lambda = -1$, the spectrum of such theory truncates to a finite number, and therefore it becomes a different description of supergravity. We found the spectrum of chiral string theories by requiring that the vertex operators are conformally invariant. In addition, we also calculate the amplitudes for arbitrary $\lambda$. The amplitudes are expressed as a product of open string amplitudes which are similar to the KLT relation. The unitarity of these theories are investigated. However, we found out that except for $\lambda = \pm 1$, all other theories are not unitary; i.e., only the supergravity and ordinary strings are unitary. Although most of the chiral strings are not physical, they still serve as a valuable tool in studying the relation between particle theories and string theories.

# Contents

# 1 Introduction

In recent years, there are efforts to understand the particle scattering amplitudes in string theory-like settings. One such effort is a method proposed by Cachazo, He and Yuan (The CHY model) [1–3], in which the massless scattering amplitude in arbitrary dimensions is expressed as a string-like integral. Specifically, the n-point tree amplitude can be written as an integral over the moduli space of $n$-punctured sphere. $\delta$ functions are inserted into the integrand to make its support lies in the solutions of the scattering equations.

In another related but independent development, Hohm, Zwiebach and Siegel found that under a certain singular gauge, the string theory spectrum apparently truncated to only contain a finite number of states. (The HSZ model) [4] The reason for this to happen is that under this gauge, the string Lagrangian become only dependent on $z$ (more precisely, the dependence of Lagrangian on $\bar{z}$ becomes infinitesimal). The HSZ model is working under the context of CFT with manifest $T$-duality symmetry; later, Siegel gives a prescription that works under ordinary bosonic string theory [5]. He also generalizes the singular gauge of the HSZ model to a continuous variation of gauges parametrized by a number $\beta$. The string Lagrangian under this gauge is given by

$$L = -\frac{1}{2}\left(\beta(\bar{\partial}X)^2 + (\partial X)(\bar{\partial}X)\right).$$

This form is achieved by taking the Lagrangian in the conformal gauge and applying the following coordinate transformation:

$$z \to z_L = \sqrt{1+\beta}z, \quad \bar{z} \to z_R = \frac{1}{\sqrt{1+\beta}}(\bar{z} - \beta z).$$

For $\beta = 0$, the transformation above is the usual conformal gauge, but in the limit of $\beta \to \infty$, it becomes the singular gauge in HSZ model. Under the mode expansion of $z_{L,R}$, the Virasoro constraints are modified in a way that result in the truncation of spectrum to infinite number. In HSZ gauge, the modified constraints correspond to a Bogoliubov transformation of right moving modes $\bar{\alpha} \to \bar{\alpha}^\dagger, \bar{\alpha}^\dagger \to -\bar{\alpha}$. This Bogoliubov transformation can be interpreted as adding a homogenous term in the string boson propagator

$$-\ln z\bar{z} \to -\ln z\bar{z} + 2\ln \bar{z} = \ln \frac{\bar{z}}{z}. \tag{1.1}$$

Although the amplitude is not computed directly in [5], it is found that under HSZ gauge, after integrating the infinitesimal dependence of $\bar{z}$ of the vertex operator, the same $\delta$ functions emerge as in the ones in CHY model.

In the paper by Huang, Siegel and Yuan, the amplitudes with the same modified boundary condition (1.1) are calculated in the conformal gauge [6]. 4-point graviton amplitude is calculated and it coincides with the one obtained from general relativity.

The KLT relation [7], in which the amplitude of a closed string is expressed as a product of open string amplitudes, is utilized in this paper. In the case of 4-point amplitudes, this relation is expressed as.

$$M_4(s,t,u) = -\pi \sin(\pi s/4) A_4(s,t) A_4(s,u),$$

where $M_4$ is the closed string amplitude, and $A_4$ is the open string amplitude. The open string amplitude is

$$A_4(s,t) = K \frac{\Gamma(-s/4)\Gamma(-t/4)}{\Gamma(1+u/4)}.$$

(In this paper, we will use the convention $\alpha'_{\text{open}} = 1/4$ and $\alpha'_{\text{closed}} = 1$. In the KLT relation, the tension for the open string amplitude needs to be replaced with the tension of the closed string.) Observe that $A_4(s,t)$ has poles of $s$ at $4n, n = -1, 0, 1 \cdots$. Thus, according to the KLT relation, the two factors of $A_4$ give double poles at $s = 4n$, but the sine factor canceled one pole out, and thus giving $M_4$ single poles at the same positions of $A_4$. These poles correspond to the exchange of the massive states of the 4-point string amplitudes.

Under the change of boundary condition (1.1), if one re-derive the 4-point amplitude in the same way as in the original paper of KLT relation, one finds a change of the second factor of $A_4$.

$$A_4 A_4 \longrightarrow A_4(s,t) \bar{A}_4(s,u) \equiv A_4(s,t) A_4(-s,-u).$$

In this case the first factor will give poles of open string amplitude at $s = 4n$ and the second gives poles at $s = -4n$. Therefore, only $s = -4, 0, 4$ have double poles in $A_4 \bar{A}_4$, while the sine factor have zeros at $s = 0, \pm 4m, m = 1 \cdots$. Thus, the remaining poles in $M_4$ are $s = 0, \pm 4$, indicating the truncation to finite number of states.

In this paper, we will generalize the boundary condition of the propagator to

$$\ln z + \lambda \ln \bar{z}, \tag{1.2}$$

where $\lambda$ is a complex number parametrizing a subspace of all the possible boundary conditions. Since the left moving and the right moving modes are treated asymmetrically, we call strings with such a boundary condition a chiral strings.

In section 2, we shall give our motivation on choosing this boundary condition. Chiral string theory is motivated by transforming the ordinary string propagator in a way analogous to the duality transformations of electric and magnetic fields.

In section 3, we shall classify all the conformally invariant states, and find all their masses. This is achieved by finding the holomorphic dimensions of various vertex operators by calculating their OPEs with respect to the energy-momentum tensors $T, \bar{T}$. Note that the definitions of $T$ and $\bar{T}$ are slightly modified to maintain the correct OPEs among themselves. This is equivalent to solving the level matching condition in [6]. While we focus on classifying the spectra of bosonic strings, the derivation for superstrings is essentially the same.

To find the amplitudes, one has to calculate the correlator of vertex operators. Unfortunately, the new propagators will introduce various cuts in the correlator and they cause ambiguities in the integral over all vertex operators positions. Therefore, we need to pay attention to the $i\varepsilon$ prescription of the propagators, in order to know how the cuts should be placed on the worldsheet. Therefore, in section 4, we re-derived the propagator by integrating the momentum modes with all the $i\varepsilon$ included. Note what while the $i\varepsilon$ prescription resolved the divergence for large momentum (and gives us the freedom to have the arbitrary parameter $\lambda$), there is still IR divergence since the string propagator is massless. To resolve this divergence, we deploy dimensional regularization, and additional constants will arise from the renormalization of the propagator. The constants can be fixed by requiring the amplitudes to be invariant under the permutation of external states ($s, t, u$ permutation invariance for 4-point amplitudes.)

In section 5, we calculated the amplitude of the chiral string using a similar method for deriving KLT relations. The final result is in the form $M_n \sim \sin(\pi k_i \cdot k_j / 2) A_n \bar{A}_n$, where $A_n, \bar{A}_n$ are the open string amplitudes, while $\bar{A}_n$ has its tension scaled by $\lambda$. We study the 4-point amplitude in detail; we found out that it has Reggie-like behavior in the Reggie limit with a Reggie slope $(1 + \lambda)/4$. In the fixed angle and large energy limit, the amplitude becomes exponentially soft for $\mathrm{Re}\lambda > -1$, while it exponentially diverges for $\mathrm{Re}\lambda < -1$. In the threshold $\mathrm{Re}\lambda = -1$, the amplitude diverges polynomially just like the supergravity amplitude. The poles in the 4-point amplitude correspond to the states which are exchanged in the scattering process. We can compare these states with the conformally invariant states we found in the previous section, and find that not all states exchanged are conformally invariant. The only cases in which all the exchanged states are conformally invariants are $\lambda = 0, \infty, \pm 1$. The unitarity of the amplitude can be checked by expanding the residues of the poles in terms of the Gegenbaur polynomial. The amplitude is unitary if and only if all the coefficients are negative. We find that only when $\lambda = \pm 1$ the theory is unitary. When $\lambda = \pm 1$, the theory become the cases covered in previous literature.

During the preparation of this paper, the authors became aware of the work done by Jusinskas [12]. In his paper, he proposed a boundary condition of the string propagator. This boundary condition is a special case of (1.2), where $\lambda = -1/N$ for some integer $N$. When evaluating the 4-point amplitude, he chose different branch cuts of the Koba-Nielson factor from us and therefore gave a different amplitude which seems to be unitary.

## 2 Boundary Condition

It is natural to generalized the Bogoliubov transformation of right moving modes in [5] into a $SL(2, \mathbb{C})$ transformation:

$$\begin{pmatrix} \bar{\alpha}'_n \\ \bar{\alpha}'^{\dagger}_n \end{pmatrix} = M \cdot \begin{pmatrix} \bar{\alpha}_n \\ \bar{\alpha}^{\dagger}_n \end{pmatrix},$$

in which $M \in SL(2, \mathbb{C})$ and $n \geq 0$. We have use the convention $ip_0 \equiv \bar{\alpha}_0$ and $x_0 \equiv \bar{\alpha}^{\dagger}_0$ here. The new vacuum is defined by the annihilation of the transformed lowering operators: $\bar{\alpha}'_n |\Omega\rangle = 0, \quad n \geq 0$.

The corresponding propagator for this boundary condition can be calculated as the following.

Letting $M = \begin{pmatrix} a & b \\ c & d \end{pmatrix}$, we have

$$\begin{aligned} G(z, w) &\equiv \sim \langle \Omega | X(z, \bar{z}) X(w, \bar{w}) | \Omega \rangle \sim -\ln(w - z) \\ &- (ad + bc) \ln(\bar{z} - \bar{w}) - (ac + bd) \ln(\bar{w}\bar{z} - 1) - bd \left[ \frac{1}{2} \ln(\bar{w}) \ln(\bar{z}) - \ln(\bar{w}\bar{z}) \right]. \end{aligned}$$

The problem of this propagator is that it is not invariant under translations, meaning that the amplitude constructed from this propagator is not generally conformally invariant. Of course, one can hope to find a translational invariant propagator by trying with Bogoliubov transformations that mix different oscillator numbers $n$, but an easier way of finding an translational invariant propagator is viewing the string propagator as a solution of 2D Guass equation with a point charge source.

Given the equation $\nabla^2 G = \delta(0)$, there are two different types of solution.
The "electric" solution:
$$G_E = \ln \rho. \quad (\nabla G_E = \hat{\rho}/\rho),$$

and the "magnetic" solution:

$$G_M = \theta. \quad (\nabla G_M = \hat{\theta}).$$

The reason we call them the electric and magnetic solutions are obvious if we plot the field lines of $\nabla G_E$ and $\nabla G_M$, they will resemble the filed lines of a point charge and a monopole. See the figures below.

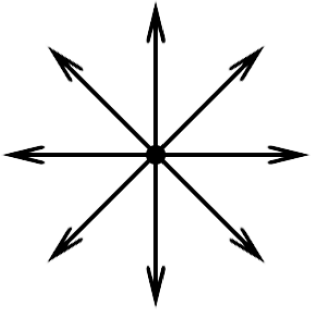

Figure 1: The field lines of $\nabla G_E$.

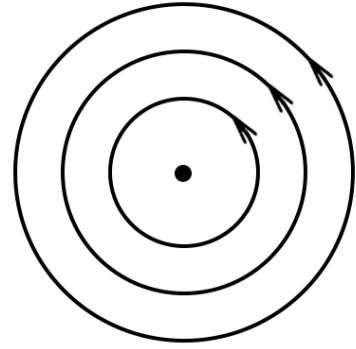

Figure 2: The field lines of $\nabla G_M$.

We have used the polar coordinate $(\rho, \theta)$ for the (Euclidean) world sheet. Written in a more familiar complex coordinate $(z, \bar{z})$, they become

$$G_E = \ln(z\bar{z}), \quad G_M = i\ln(z/\bar{z}).$$

As we can see, using the electric propagator will result in the usual string amplitudes, and it is shown in [5], using the magnetic propagator will result in a truncated amplitude with only finite particle states.

Since both $G_M$ and $G_E$ are solutions to the 2D gauss equation, it is natural to consider the linear combination of them. The chiral string propagator is therefore defined as

$$G_C(z, \bar{z}) = \ln z + \lambda \ln \bar{z},$$

$z$ represents the contribution of the left moving modes and $\bar{z}$ represents the contributions of right moving modes. $\lambda \in \mathbb{C} \cup \{\infty\}$ and $\lambda \to \infty$ is the limit where only right moving modes exist.

This linear combination of "electric" and "magnetic" propagators has a natural analogy to the EM duality of 4D Maxwell equations. The EM duality states that if $E$ and $B$ are solutions to the Maxwell equations, one can rotate the solutions by a phase angle $\theta$

$$E' + iB' = e^{i\theta}(E + iB).$$

$E'$ and $B'$ are still the solutions of Maxwell equations. If we consider the identification

$$E \to G_E, \quad B \to G_M, \quad D \equiv E + iB \to G_E + iG_M = 2\ln z.$$

A duality rotation will rotate the normal string propagator ($G_E$) to the chiral propagator:

$$E' = \text{Re}(e^{i\theta}D) \quad \Rightarrow \quad G_C = \text{Re}(e^{i\theta}2\ln z) = e^{i\theta}\ln z + e^{-i\theta}\ln\bar{z}.$$

(Remember that we can always absorb the constant proportional to $\ln z$ into the string tension, making $\lambda = e^{-2i\theta}$.)

The OPE between $X$ and itself for chiral strings is therefore modified to be

$$\langle X(z, \bar{z})X(0) \rangle \sim -\frac{1}{2}(\ln z + \lambda \ln \bar{z}).$$

we have set the string tension to be $\alpha' = 1$ for closed strings. OPEs of the ghosts (and fermions if we are considering superstrings) can be derived in a similar manner:

$$\langle b(z)c(0) \rangle \sim \frac{1}{z}, \quad \langle \bar{b}(\bar{z})\bar{c}(0) \rangle \sim \lambda \frac{1}{\bar{z}}.$$

The stress tensors of chiral string theory are slightly modified.

$$T = -\partial X \partial X, \quad \bar{T} = -\frac{1}{\lambda}\bar{\partial}X\bar{\partial}X.$$

(The ghost contribution is ignored.) We can use the operators above as the definitions of stress tensors since they obey the usual OPE that defines the stress tensors.

$$TT \sim \frac{2T}{z^2} + \frac{\partial T}{z} + \frac{c}{z^4}, \quad \bar{T}\bar{T} \sim \frac{2\bar{T}}{\bar{z}^2} + \frac{\bar{\partial}\bar{T}}{\bar{z}} + \frac{c}{\bar{z}^4},$$

where $c$ is the central charge. As one can see, the pattern is very clear. The chiral string theory is only different from the ordinary string theory by replacing the string tension of the anti-holomorphic sector with $\lambda$. The previous statement holds true even when we are calculating the amplitudes, or dealing with other string theories such as superstrings.

## 3 Spectra

One way to find the spectrum of the chiral string theory is to find the valid vertex operators for them. Consider the integrated vertex operator

$$U = \int dz d\bar{z} V(z, \bar{z}),$$

the corresponding unintegrated vertex operator can be constructed

$$W(z, \bar{z}) = c\bar{c}V(z, \bar{z}).$$

In either case, in order for the vertex operator to be conformally invariant, one needs $V$ to have a holomorphic dimension $(1, 1)$: i.e. $V$ has the following OPEs

$$T(z)V(0) \sim \frac{hV(0)}{z^2} + \frac{\partial V(0)}{z}, \quad h = 1$$

$$\bar{T}(z)V(0) \sim \frac{\bar{h}V(0)}{\bar{z}^2} + \frac{\bar{\partial}V(0)}{\bar{z}}, \quad \bar{h} = 1.$$

For chiral string theory, the general vertex operator is of the form

$$V^{(n,m)} \equiv V^{(\mu_1\cdots\mu_n)(\nu_1\cdots\nu_m)}(z,\bar{z};k) = \prod_i^n \partial X^{\mu_i} \prod_j^m \bar{\partial} X^{\nu_j} e^{ikX}.$$

The holomorphic dimension for such an operator can be found by calculating the OPE of $V$ with $T, \bar{T}$, using the new $\langle XX \rangle$ propagator discussed in the previous section. One can find that the holomorphic dimension of $V$ is

$$(h, \bar{h}) = \left(n + \frac{k^2}{4}, m + \frac{\lambda k^2}{4}\right).$$

By requiring that $h = 1, \bar{h} = 1$, one can find the valid states $(n, m)$ and their corresponding mass $M^2 = -k^2$.

Observe that for any $\lambda$, $(n, m) = (1, 1)$ and $M^2 = 0$ is always a solution. Namely, the massless sector is always valid. The second observation is that $\lambda = \frac{m-1}{n-1}$. Therefore, if $\lambda$ is irrational, the massless sector is the only set of states that are conformally invariant. On the other hand, if $\lambda = p/q \geq 0$, there will be an infinite tower of states just like ordinary string theory:

$$(n, m) = (kq + 1, kp + 1) \quad M^2 = 4kq = \frac{4kp}{\lambda}.$$

$k$ needs to be integers that make $m, n$ non-negative integers. Clearly, this formula also holds when $\lambda$ is a negative rational number. But since $m, n$ are positive, there will be only finite numbers (less then three) states that are valid.

We have chosen to scale the tension of the right-moving sector with $\lambda$ instead of the left-moving sector. Choosing to scale the left-moving sector will not result in new theories, since it is just a transformation of parameters. $\lambda \to 1/\lambda$. In the case that $\lambda = 0$ (Pure left moving strings.), this transformation needs to be treated as a limit $\lambda \to \infty, \alpha' \to 0$, such that $\lambda \alpha'$ equals 1.

The complete classification of all the conformal states for all $\lambda$ is given in appendix A. Note that this is just a classification of all the possible conformal states. It does not necessarily mean that a chiral string for a particular $\lambda$ will only have these states, even though only having conformally invariant states is necessary for the theories to be free of Weyl anomalies. In section 5.1, we will find all the states in chiral strings by finding the poles in their amplitude. By comparing the states to the table above, we can give constraints on the possible values of $\lambda$ that give consistent theories.

## 3.1 Superstrings

One can in principle find the conformally invariant spectra for (chiral) superstrings by classifying all the possible vertex operators of holomorphic dimension $(1, 1)$. While it is necessary to find the vertex operators to calculate the amplitude, one doesn't have to find the vertex operators if one is only interested in the spectra.

Notice that the conditions for the general vertex operators $V^{(n,m)}$ to have dimension $(1,1)$ are

$$\frac{k^2}{4} + n - 1 = 0, \quad \frac{\lambda k^2}{4} + m - 1 = 0$$

Observe that these conditions are equivalent to

$$L_0^{(B)}|\text{phys}\rangle = 0, \quad \bar{L}_0^{(B)}|\text{phys}\rangle = 0$$

$L_0^{(B)}, \bar{L}_0^{(B)}$ are the 0th order generators of Virasoro algebra for bosonic strings.

$$L_0^{(B)} = \frac{p^2}{4} + \sum_{n>0} a_{-n}^i a_n^i - 1, \quad \bar{L}_0^{(B)} = \frac{\lambda p^2}{4} + \sum_{n>0} \tilde{a}_{-n}^i \tilde{a}_n^i - 1$$

It is easy to generalize this condition to include the fermionic modes. For convenience, we will use Green-Schwarz formalism in the light-cone gauge. The fermionic modes are $S_n^a, \tilde{S}_n^a$, in which the index $a$ is the 8D spinor of the little group. For concreteness, we have chosen the indices of both left-moving and right-moving modes to have the same charity; i.e. type IIA string theory. One can also choose the right moving modes to have spinor index with opposite chirality; i.e. type IIB string theory. The classification is essentially the same for both type IIA and B. The 0th order generators of Virasoro algebra including the fermionic modes are

$$L_0 = \frac{p^2}{4} + \sum_{n=0}^{\infty} a_{-n}^i a_n^i + \sum_{n=0}^{\infty} S_{-n}^a S_n^a$$
$$\bar{L}_0 = \frac{\lambda p^2}{4} + \sum_{n=0}^{\infty} \tilde{a}_{-n}^i \tilde{a}_n^i + \sum_n \tilde{S}_{-n}^a \tilde{S}_n^a.$$

Therefore, the condition $L_0|\psi\rangle = \bar{L}_0|\psi\rangle = 0$ translate into

$$\begin{cases} \frac{M^2}{4} = N \\ \frac{\lambda M^2}{4} = \tilde{N} \end{cases}$$

In which $N, \tilde{N} = 0, 1 \cdots$ are the levels of left-moving and right moving modes. (i.e., the eigenvalues of $\sum a_{-n} a_n + S_{-n}^a S_n^a$ and its right-moving counterpart).

The classification of conformal states can be found by solving the equations above. As one can see, the classification is only dependent on the levels of the states and not on the individual representations within the levels.

For $\lambda < 0$ or $\lambda$ is irrational, the only conformal states are the massless states, while for $0 < \lambda = p/q$, there is always an infinite tower of conformal states

$$(N, \tilde{N}) = (qn, pn) \text{ where } n = 0, 1 \cdots \quad M^2 = 4qn = 4pn/\lambda.$$

For $\lambda = 0$, the conformal states are in the form

$$(\text{massless})_L \otimes (\text{open string})_R.$$

For the case when $\lambda \to \infty, \alpha'\lambda = 1$, the conformal states are

$$(\text{open string})_L \otimes (\text{massless})_R.$$

## 4 Propagators

In this section, we are going to re-derive the propagator that is presented at the beginning of section 2. The difference is that we are going to pay attention to the $i\varepsilon$ prescription of the propagator, so we can determine how the cuts will be placed on the world sheet. The position of the cuts are crucial since that in the later sections when we try to calculate the amplitude, we need to derive the chiral string version of KLT relation. Then, we can write our final chiral string amplitude as products of open-string amplitudes. To derive the KLT relation, we need to analytically continue the integration region from $(z, \bar{z})$ to contours on two independent complex plane $(\eta, \xi)$. The different choices of cut will result in different phase factors (of the integrand of the amplitude) in different integration regions, and therefore determine the outcome of the modified KLT relation.

In the rest of the paper, we will work on the Lorentzian world sheet instead of the Euclidean world sheet. We will use $\sigma_\pm = \sigma \pm t$ instead of $(z, \bar{z})$. Although the complex coordinate is useful when analyzing the OPE and spectra of string theories, the derivation of propagator by integrating the momenta modes) will be easier if the derivation is performed in $\sigma_\pm$.

One subtlety is that the $i\varepsilon$ prescription actually dose not completely fix the boundary condition of the propagator. The boundary condition is not fixed because there is IR divergence for the massless propagator and one needs further regularization to get finite value of the propagator. The regularization will introduce ambiguity in the form of additional terms such as $c_+ \theta(\sigma_+)$ added to the propagator. The constant $c_+$ is undetermined. Interestingly, the $c_+$ (and other constants) can be fixed by requiring the final amplitude to be invariant under permutations of external states ($s, t, u$ invariant).

## 4.1 Derivation

We will start with the massive 2D scalar propagator (we will take $M^2 \to 0$ later.)

$$G(\sigma_+, \sigma_-) = \int \frac{dk^+ dk^-}{(2\pi)^2} \frac{e^{i(k^+\sigma_+ + k^-\sigma_-)}}{-k^+k^- + M^2 + i\epsilon}. \tag{4.1}$$

The integrand has one pole at $k^- = \frac{M^2 + i\epsilon}{k^+}$. The contour of $k^-$ is the standard semi-circle. It goes over the positive imaginary axis if $\sigma_- > 0$, and goes over negative imaginary axis if $\sigma_- < 0$. For $\sigma_- > 0$, the poles are in the contour if $k^+ > 0$, while for $\sigma_- < 0$, the poles are in the contour if $k^+ < 0$.

Integrating $k^-$ out, and after some algebraic manipulation, we get

$$\begin{aligned}
G(\sigma_+, \sigma_-) &= \frac{1}{2\pi i} \int_0^\infty \frac{dk^+}{k^+} e^{i[k^+\sigma_+ + (M^2 + i\epsilon)\sigma_-/k^+]} \theta(-\sigma_-) \\
&+ \frac{1}{2\pi i} \int_0^\infty \frac{dk^+}{k^+} e^{-i[k^+\sigma_+ + (M^2 + i\epsilon)\sigma_-/k^+]} \theta(\sigma_-) \\
&\equiv G_+(\sigma_+, \sigma_-)\theta(-\sigma_-) + G_-(\sigma_+, \sigma_-)\theta(\sigma_-)
\end{aligned}$$

If we instead integrate $k^+$ first, we will get the same expression but with $\sigma_\pm$ exchanged. The exchange of the $\sigma_\pm$ can also be achieved by making the coordinate transformation (and with some careful analytic continuation):

$$\begin{cases} k^- = \frac{M^2 + i\epsilon}{k^+}. & \sigma_+\sigma_- > 0 \\ k^- = -\frac{M^2 + i\epsilon}{k^+}. & \sigma_+\sigma_- < 0 \end{cases}$$

This kind of coordinate transformation is expected since the original expression of the propagator is symmetric in $\sigma_\pm$. If we take the massless limit $M^2 \to 0$, $G$ becomes

$$G(\sigma_+, \sigma_-) \to G_+(\sigma_+)\theta(-\sigma_-) + G_-(\sigma_+)\theta(\sigma_-),$$

where

$$\begin{aligned}
G_+(\sigma_+) &\equiv \int_0^\infty \frac{dk}{k} e^{ik(\sigma_+ + i\epsilon)} \\
G_-(\sigma_+) &\equiv \int_0^\infty \frac{dk}{k} e^{-ik(\sigma_+ - i\epsilon)}.
\end{aligned}$$

We have absorbed $|\sigma_-|/k^2$ into $\epsilon$. What we got is the pure left-moving propagator. If we do the coordinate transformation in $k$ (or integrate $k^-$ first) and take the massless limit, we will get the same expression except that $\sigma_\pm$ are swapped and the outcome will be the pure right-moving propagator. To get the mixed chiral propagator, we can just split the massive propagator into a part which is proportional to $\lambda/(1+\lambda)$ and another proportional to $1/(1+\lambda)$, making a coordinate transformation in the first term, then taking the massless limit.

It seems that here is a contradiction. How can the original massive propagator, which is symmetric in $\sigma_\pm$, give rise to purely left-moving and purely right-moving propagator? The reason is that while the $i\varepsilon$ prescription avoids the divergence of $G_\pm$ in high energy, inferred divergence also occurs at the massless limit. Indeed, the current integral for $G_\pm$ is divergent. In order for the integral to make sense, we need to find the analytic continuation of it. The left-moving and right-moving (and the mixed) propagator all correspond to different analytic continuations of the original form 4.1.

We shall apply a regularization scheme to get the finite value of the propagator. This scheme is similar to the conventional dimensional regularization in which the integrand becomes

$$\frac{dk}{k} \to \frac{dk}{k^{1+\delta}}.$$

We shall calculate $G_+(\sigma_+)$:

$$\begin{aligned}
G_+(\sigma_+) &= \frac{1}{2\pi i}\int_0^\infty \frac{dk}{k^{1+\delta}} e^{ik(\sigma_+ + i\varepsilon)} \\
&= \frac{[-i(\sigma_+ + i\varepsilon)]^\delta}{2\pi i}\int_0^{\infty e^{-i(\pi/2-\varepsilon)}} s^{-(1+\delta)} e^{-s} ds \\
&= \frac{[i(\sigma_+ + i\varepsilon)]^\delta}{2\pi i}\Gamma(-\delta) \\
&= \frac{i}{2\pi}\left[\frac{1}{\delta} + \ln(\sigma_+ + i\varepsilon) + O(1)\right].
\end{aligned}$$

From the first equality to the second equality, we substitute $s = -ik(\sigma_+ + i\varepsilon)$. From the second equality to the third equality, we deform the integration contour from $[0,\ \infty e^{-i(\pi/2-\varepsilon)}]$ to $[0, \infty]$. Therefore, the integral is just a gamma function. This deformation can be done because the arc between these two lines vanishes at infinity. The third to the fourth equality is just an expansion in $\delta$. The $1/\delta$ term should be dropped, leading to an undetermined additive constant in the final result.

Similar arguments apply when evaluating $G_-(\sigma_+)$. Our final result for the pure left-moving propagator is

$$G_L(\sigma_+,\sigma_-) = \frac{i}{2\pi}\left[\ln(\sigma_+ + i\varepsilon) + c_+\right]\theta(-\sigma_-) + \frac{i}{2\pi}\left[\ln(\sigma_+ - i\varepsilon) + c_-\right]\theta(\sigma_-),$$

and the pure right moving propagator is

$$G_R(\sigma_+,\sigma_-) = \frac{i}{2\pi}\left[\ln(\sigma_- + i\varepsilon) + b_+\right]\theta(-\sigma_+) + \frac{i}{2\pi}\left[\ln(\sigma_- - i\varepsilon) + b_-\right]\theta(\sigma_+).$$

$c_\pm, b_\pm$ are undetermined constants from renormalization. These constants can be interpreted as additional homogenous solutions $\theta(\pm\sigma_\pm)$ of the 2D wave equation. These terms correspond to a part of boundary condition that is not fixed by $i\varepsilon$ prescription.

If one fix $\sigma_\pm$ to be real and non-zero, $G_R, G_L$ is can simplified to

$$G_L = \frac{i}{2\pi}(\ln|\sigma_+| + \Delta_L), \quad G_R = \frac{i}{2\pi}(\ln|\sigma_-| + \Delta_R).$$

$\Delta_R, \Delta_L$ are phase factors originated from the cuts. They are

$$\Delta_L = \begin{cases} 0 & \sigma_+ > 0 \\ i\pi + c_+ & \sigma_- > 0, \sigma_+ < 0 \\ -i\pi + c_- & \sigma_- < 0, \sigma_+ < 0 \end{cases}, \quad \Delta_R = \begin{cases} 0 & \sigma_- > 0 \\ i\pi + b_+ & \sigma_+ > 0, \sigma_- < 0 \\ -i\pi + b_- & \sigma_+ < 0, \sigma_- < 0 \end{cases}.$$

The chiral propagator is $G_C = G_L + \lambda G_R$. The 4-point amplitude calculated for general values of $b_\pm, c_\pm$ will not be $s, t, u$ invariant.

The condition that the amplitude needs to be $stu$ invariant will completely fix the constant. Two solutions are found:

$$c_+ = 0, \ c_- = i\pi, \ b_+ = (1-1/\lambda)i\pi, \ b_- = i\pi, \tag{4.2}$$

and

$$c_+ = (1-\lambda)i\pi, \ c_- = i\pi, \ b_+ = 0, \ b_- = i\pi. \tag{4.3}$$

The second set of solutions is just a redefinition $\lambda \to 1/\lambda$ as well as swapping left moving sector and right moving sector.

The chiral propagator can therefore be written as

$$G_C = \frac{i}{2\pi}(\ln|\sigma_+| + \lambda \ln|\sigma_-| + \Delta_C)$$

The phase factor $\Delta_C$ can be expressed in a compact form:

$$\Delta_C = i\pi\theta(\sigma_+\sigma_-) \quad \text{(First solutions)} \quad \Delta_C = i\lambda\pi\theta(\sigma_+\sigma_-) \quad \text{(Second solutions)}$$

## 5 Amplitudes

We shall first focus on the formulation of $n$-point graviton tree amplitude in the chiral string and later explicitly calculate the 4-point amplitude. We will find out that the 4-point amplitude is natural to be represented as a kind of Kawai-Lewellen-Tye (KLT) type factorization.

Recall that the string tree amplitude is a correlation function of 3 unintegrated operator and $n-3$ integrated operator:

$$M_n(k_1, \cdots k_n) = \left\langle W(\hat{\sigma}_1; k_1)W(\hat{\sigma}_2; k_2)W(\hat{\sigma}_3; k_3) \prod_{i \leq 3} U(k_i) \right\rangle_{S^2}.$$

$W(\hat{\sigma}_1, k_1)$ and $U(k_i)$ are the graviton vertex operators:

$$W(\sigma; k) = c_+ c_- V(\sigma; k) \quad U(k) = \int d^2 V(\sigma; k) \quad V(\sigma; k) = \zeta_{\mu\nu} \partial_+ X^\mu \partial_- X^\nu e^{ik \cdot X}.$$

$\zeta_{\mu\nu}$ are the polarization tensors. To facilitate the calculation, we assume that the momenta of all the particles lie inside a 4-dimensional subspace, so in the transverse traceless gauge, one can write $\zeta_{\mu\nu} = \zeta_\mu^+ \zeta_\nu^-$, where $\zeta^\pm$ are transverse (photon) polarization vectors. A generalization to higher dimensions is trivial. Another trick we will use is exponentiating $\zeta^\pm$ in the vertex operators and extracting the linear terms after the path integral is evaluated:

$$\zeta^+ \zeta^- \partial_+ X \partial_- X e^{ik \cdot X} \to e^{ik \cdot X + \zeta^+ \partial_+ X + \zeta^- \partial_- X}.$$

After evaluating the path integral, we get

$$M_n \stackrel{\text{linear}}{=} C\delta\left(\sum k_i\right) \Delta_+ \Delta_- \int \prod_{k \geq 3} d^2 \sigma_k \exp -\frac{i}{2} \left[ \sum_{i \neq j} k_i k_j G(\sigma_{ij}) + k_i \zeta_j^+ \partial_+ G(\sigma_{ij}) + \right.$$
$$\left. + k_i \zeta_j^- \partial_- G(\sigma_{ij}) + \zeta_i^+ \zeta_j^+ \partial_+^2 G(\sigma_{ij}) + \zeta_i^- \zeta_j^- \partial_-^2 G(\sigma_{ij}) \right]$$

The notation $\stackrel{\text{linear}}{=}$ means that the RHS needs to be expanded in terms of $\zeta$ and only keep the terms proportional to $\zeta_1^+ \zeta_1^- \cdots \zeta_n^+ \zeta_n^-$. To get a gauge independent form, one just has to replace each pair of $\zeta_i^\pm$ with the original polarization tensors $\zeta_{\mu\nu}$. The constant $C$ is a contribution from the closed string coupling (i.e., vacuum expectation value of dilaton) and the functional determinates of the laplacians of the bosons and the ghosts. They are naively infinite and therefore need regularization. Notice that there is no need to calculate $C$ explicitly. Since $C$ does not depend on the number vertex operators and the boundary condition of the propagator ($\lambda$), one can find its value by comparing the residue of the 4-point amplitude and the 3-point amplitude in ordinary string theory. We are going to find its value later.

$\Delta_\pm$ is just the correlator of the ghosts:

$$\Delta_+ = (\hat{\sigma}_{+1} - \hat{\sigma}_{+2})(\hat{\sigma}_{+2} - \hat{\sigma}_{+3})(\hat{\sigma}_{+3} - \hat{\sigma}_{+1}), \quad \Delta_- = \lambda(\hat{\sigma}_{-1} - \hat{\sigma}_{-2})(\hat{\sigma}_{-2} - \hat{\sigma}_{-3})(\hat{\sigma}_{-3} - \hat{\sigma}_{-1})$$

To further evaluate the amplitude, we substitute the chiral propagator with the expression we derived in the previous section

$$G(\sigma_{ij}) = \frac{i}{2} \left( \ln |\sigma_{ij+}| + \lambda \ln |\sigma_{ij-}| + i\pi \theta(\sigma_{ij+} \sigma_{ij-}) \right).$$

The choices of $b_\pm, c_\pm$ are given by 4.2. The n-point amplitude becomes

$$M_n \stackrel{\text{linear}}{=} C\delta\left(\sum k_i\right) \Delta_+ \int_{-\infty}^{\infty} \prod_{3 \leq k}^{n} d\sigma_{k+} \prod_{i<j} |\sigma_{ij+}|^{\alpha' k_i k_j / 2} \exp\left[ 4 \sum_{i \neq j} \frac{\zeta_i^+ k_j}{\sigma_{ij+}} - \frac{2\zeta_i^+ \zeta_j^+}{\sigma_{ij+}^2} \right]$$
$$\cdot \Delta_- \int_{-\infty}^{\infty} \prod_{3 \leq k}^{n} d\sigma_{k-} \prod_{i<j} |\sigma_{ij-}|^{\lambda \alpha' k_i k_j / 2} \exp\left[ \frac{\lambda}{4} \sum_{i \neq j} \frac{\zeta_i^- k_j}{\sigma_{ij-}} - \frac{2\zeta_i^- \zeta_j^-}{\sigma_{ij-}^2} \right] e^{i\pi A(\sigma_1, \cdots \sigma_n)}$$

Note that $\partial_\pm G$ and $\partial_\pm^2 G$ will give terms proportional to $\delta(\sigma)$ and $\delta'(\sigma)$. They will give additional contact terms that are evaluated at $\sigma_i = \sigma_j$ in the amplitude, but by the nature of Koba-Nielson factor, these terms always vanish. The phase $A(\sigma_1, \cdots \sigma_n)$, which is the result of additional constants $\Delta_c$ in the propagators, equals to

$$A(\sigma_{1+}, \sigma_{1-}, \cdots \sigma_{n-}) = \frac{1}{2} \sum_{i<j} k_i k_j \theta(\sigma_{ij+} \sigma_{ij-}).$$

Observe that written in this form, the chiral string amplitude is almost a product of two open string amplitude: one with a tension 1 and the other with tension $\lambda$. However, the two integrals are twisted by the phase factor $e^{i\pi A}$.

We shall now focus on the case when $n = 4$ and take the position of unintegrated vertex operators to $\hat{\sigma}_1 = 0, \hat{\sigma}_2 = 1, \hat{\sigma}_3 = \infty$. The only coordinate one needs to integrate is $\sigma_4^\pm$.

First, consider the open strings amplitude before integrating over the vertex operator positions, we can divide the integration region into $[-\infty, 0)$, $[0, 1)$, and $[1, \infty)$. These integrations regions correspond to the $s, t$ and $u$ channel of open string amplitudes.

$$\int_{-\infty}^{\infty} |\sigma_4^+|^{-\alpha' u/4} |1 - \sigma_4^+|^{-t/4} C(\zeta^\pm, k_i; \sigma_4^+) d\sigma_4^+ = A_4(u, s) + A_4(u, t) + A_4(s, t),$$

where $C(\xi^\pm, k_i; \sigma_4^+)$ is some complicated polynomial in terms of $\sigma_4^+, \zeta^\pm$ and $k_i$. It gives the kinematic factor $K$ in open string amplitude.

$$A_4(u, t) = K \frac{\Gamma(-u/4)\Gamma(-t/4)}{\Gamma(1 + s/4)}, \text{ etc.}$$

$K \equiv K_{\alpha\beta\gamma\delta} \zeta^\alpha \zeta^\beta \zeta^\gamma \zeta^\delta$ is some complicated contractions between the momenta and polarization vectors. The explicit forms of $K$ are given in appendix A. The only thing worths noting is that $K$ is invariant under permutations of 4 external states and it contains pole at $s, t, u = -4$, which indicates an exchange of tachyon.

Now, we can see that we can divide the integration region of $M_4$ into 9 parts. As we can choose $\sigma_4^\pm$ to be either in one of three intervals $[-\infty, 0), [0, 1), [1, \infty)$ and since that the phase factor is constant in each of the integration region, the integral factors out nicely into the form $A_R A_L e^{i\pi A}$.

$A_R, A_L$ are both open string amplitudes listed above. The only difference is that $A_R$ has $\lambda$ as tension instead of 1. Collecting all nine terms and after some algebraic manipulation, one arrives at the final form of 4-point chiral string amplitude

$$M_4 = \frac{1}{3} \left( \sin(\pi u/4) A_R(s, u) A_L(u, t) + \sin(\pi s/4) A_R(t, s) A_L(s, u) + \sin(\pi t/4) A_R(u, t) A_L(t, s) \right).$$

We have omitted the $\delta(\sum k_i)$ factor in this expression. Also, the 1/3 factor comes from fixing the constant $C$ using the relation between the s-channel residue of the tachyon and 3-point amplitudes.

$$\operatorname*{Res}_{s=-4} M_4(1,2,3,4) = M_3^*(1,2,p) M_3(-p,3,4).$$

$p$ represents the momentum of the tachyon exchange. Since $C$ is independent of $\lambda$, it is sufficient to solve $C$ when $M_4, M_3$ are ordinary string amplitudes and the result will apply in the general case.

$M_4$ is $stu$ invariant. The invariance becomes obvious if we write it in terms of Gamma functions

$$\begin{aligned} M_4 &= \frac{1}{3} K_R K_L \left[ \frac{\Gamma(-\lambda s/4)\Gamma(-\lambda u/4)\Gamma(-t/4)}{\Gamma(1+s/4)\Gamma(1+u/4)\Gamma(1+\lambda t/4)} + (u \leftrightarrow t) + (s \leftrightarrow t) \right] \\ &\equiv K_R K_L C(\lambda; s, t, u) \end{aligned} \quad (5.1)$$

$K_R, K_L$ are the left-moving and right moving kinematic factors and just as $A_R, A_L$, the tension of right moving kinematic factor is scaled by $\lambda$. Note that in the derivation above, we used the boundary condition 4.2 for the propagators. If we use 4.3 instead, we will get a different result. However, the different result is essentially just swapping the role of left-moving and right-moving mode (i.e., $\lambda \to 1/\lambda$). Therefore, this equation covers both scenarios.

While $M_4$ we derived above is the 4-graviton amplitude for bosonic string theory, it can be easily generalized into other cases. For example, one can combine all the amplitudes of the massless sector into a unified form (Including the graviton, the 2-form and the dilaton).

$$M_{4,\text{massless}} = e^1_{\alpha\mu} e^2_{\beta\nu} e^3_{\gamma\rho} e^4_{\delta\sigma} K_R^{\alpha\beta\gamma\delta} K_L^{\mu\nu\rho\sigma} C(\lambda; s, t, u).$$

$K_{R,L}^{\alpha\beta\gamma\delta}$ are $K_{R,L}$, except that the polarization vectors are stripped. The indices of $K_{R,L}^{\alpha\beta\gamma\delta}$ are contracted to the indices of polarization tensors $e_{\mu\nu}$ instead. They are the combined polarization tensors of all massless sectors of bosonic string theory.

$$e_{\mu\nu} = \underbrace{e_{(\mu\nu)}}_{\text{graviton}} + \underbrace{e_{[\mu\nu]}}_{\text{2-form}} + \underbrace{\frac{1}{d} e^\mu_\mu}_{\text{dilaton}}$$

The amplitude for the superstring is exactly the same form, but $K_{L,R}$ are replaced by $K_{0\,L,R}$, which is defined in the appendix B.

## 5.1 Poles

By studying the poles the 4-point amplitude, one can get a picture of the spectrum possessed by the chiral string theory. By $stu$ symmetry, it is sufficient to find just the position of poles $s$. Looking at the full amplitudes in terms of gamma functions (equation 5.1), one can see that places where poles can arise are where the gamma functions in the numerator are taking non-positive integers:

$$s_{\text{pole}} = 4n, \quad \frac{4n}{\lambda}, \quad n = -1, 0 \cdots$$

On the other hand, one may also need to consider the zeros of the amplitude, since they might cancel the poles we just describe above. The zeros only occur when the gamma functions in the denominator are taking non-positive integers.

$$s_{\text{zero}} = -4(n+1) \quad \frac{4(n+1)}{\lambda}, \quad n = 0 \cdots.$$

The poles and zeros of the amplitude for generic $\lambda$ are summarized in figure 3. The dots represent poles, and the x-marks represents zeros. If the dots coincide with the x-marks, the poles and zeros cancel each other. However, there is a slight complication: since the amplitude 5.1 contains three terms, the zero and poles will only cancel each other if and only if they are in the same term. Therefore, the dots and x-marks in the figure are color coded. The first and second term have the same poles and zeros and they are represented in red, while the zeros and poles of third term are represented in blue.

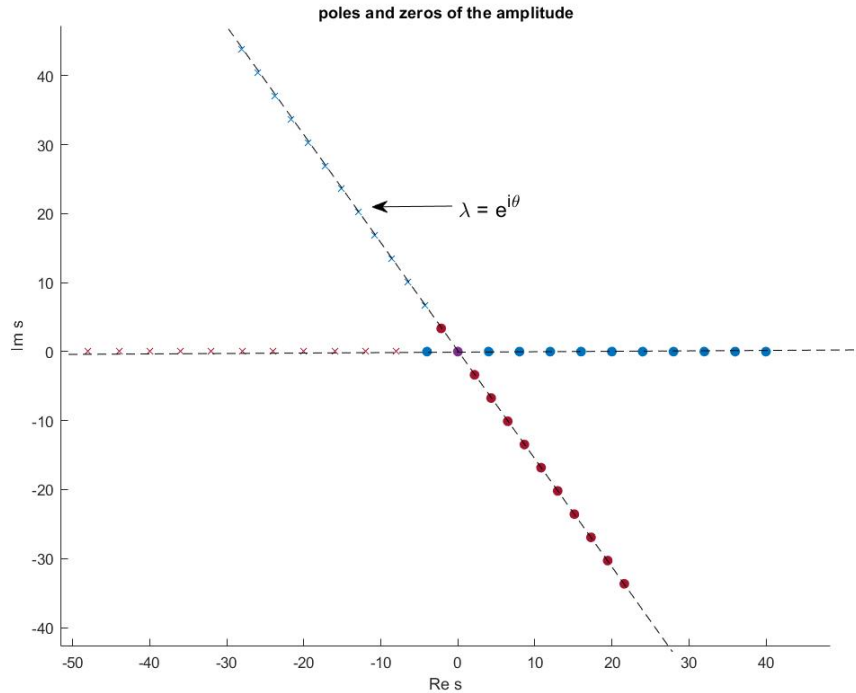

Figure 3: This figure represents the structure of the poles and zeros of the chiral string amplitude 5.1. The dots represents poles and the x-marks represents zeros. The colors indicate the terms where zeros and poles are contained. Red means they are contained in the first and the second term, while blue indicates they are contained in the third term. Purple means that the pole or zero is contained in all three terms.

From figure 3, we can see that the dots and x-marks of the same color coincide only when $\lambda = -1$. i.e., Almost all poles and zeros cancel each other, leaving only finite (three) states in the theory.

The poles in the amplitude represent a particle exchange. One can directly read the masses of the particles by the position of the pole. Comparing masses with the classification of conformal states we found in the previous section, we can see that for general $\lambda$, there always are some non-conformal states exchanged through the s-channel.

There are exactly four cases in which all the exchanged states are conformally invariant: $\lambda = 1$ (ordinary strings ), $\lambda = -1$ (field theory), $\lambda = 0$, and $\lambda \to \infty$, $\lambda \alpha' = 1$. For the latter two cases, the amplitudes are in the form of (Open string) $\otimes$ (Massless). For $\lambda = 0$, we have

$$M_{4,\lambda \to 0} = \frac{16}{3\lambda^2} K_R K_L \left[ \frac{1}{st} \frac{\Gamma(-u/4)}{\Gamma(1+t/4)\Gamma(1+u/4)} + (s \leftrightarrow t) + (s \leftrightarrow u) \right]$$

Notice that $K_R$ is proportional to $\lambda^2$, canceling the $1/\lambda^2$ factor in the front; for $\lambda \to \infty$

$$M_{4,\lambda \to \infty}/\alpha' = \frac{4}{3} K_R K_L \left[ \frac{1}{s} \frac{\Gamma(-t/4)\Gamma(-u/4)}{\Gamma(1+s/4)} + (s \leftrightarrow t) + (s \leftrightarrow u) \right].$$

## 5.2 Energy limits

Under the Regge limits $s \to -\infty + i\epsilon$ and $t << |s|$ f, the 4-point massless amplitude becomes

$$M_4 \sim \left[ f_1(t) e^{i\pi(1-\lambda)|s|/4} + f_2(t) \right] |s|^{(1+\lambda)t/4},$$

where $f_1(t), f_2(t)$ are complicated functions of $t$. This amplitude has a Regge-like behaviour with a regge slope $(1+\lambda)/4$. The only difference between this amplitude and the one from the usual Regge theory is the extra phase factor $\pi(1-\lambda)|s|/4$. The slope of the Regge slope is consistent with the classification of the conformal states. In the spectra, the maximum spin in a given level $(N, \tilde{N})$ is $J_{\max} = N + \tilde{N}$. Since the conformally invariant states obey $N = \lambda \tilde{N}$ and $M^2 = 4N$, we indeed have $J_{\max} = M^2(1+\lambda)/4$.

We can also consider the fixed angle and large energy limit, where $s \to \infty + i\epsilon$, while $s/t = -(1-\cos\theta)/2$ is constant. In this limit, we get

$$M_4 \sim s e^{(1+\lambda)f(\theta)s}, \quad f(\theta) = \left[ \frac{1+\cos\theta}{2} \ln\left(\frac{1+\cos\theta}{2}\right) + \frac{1-\cos\theta}{2} \ln\left(\frac{1-\cos\theta}{2}\right) \right].$$

Since $f(\theta) < 0$, the amplitude is exponentially soft if $\text{Re}\lambda > -1$ and diverges exponentially if $\text{Re}\lambda < -1$; at the critical value $\text{Re}\lambda = -1$ the amplitude diverges in the order of $O(s)$ just like field theory.

## 5.3 Unitarity

We can find the necessary condition for the theory to be unitary by studying the residues of its amplitudes. Observe that the 4-point tree amplitude in $D$ dimensions, with massless scalars as external states and exchanging massive spin $l$ through the s channel is uniquely determined:

$$M_{l,\text{4-scalars}} = -g^2 \frac{G_l^{(D)}(\cos\theta)}{s - m^2}.$$

$\theta$ is the scattering angle in the center of mass frame. $G_l^{(D)}(x)$ is the Gegenbauer polynomial in $D$ dimensions. The Gengenbuer polynomials are the generalizations of Legendre polynomials. They are defined by the expansion

$$(1 - 2xt + t^2)^{-(D-3)/2} = \sum_{l=0}^{\infty} t^l G_l^{(D)}(x).$$

Suppose that there is some general massless 4-scalar amplitude in $D$ dimension denoted as $M(s,t)$. This amplitude has a s-channel poles at $m^2$ and its residue has a decomposition

$$\operatorname*{Res}_{s=m^2} M(s,t) = -\sum_{l=0}^{\infty} g_l^2 G_l^{(D)}(x),$$

where $x \equiv \cos\theta$. This decomposition is always possible since $G_l^{(D)}$ are orthogonal polynomials. However, since each term is interpreted as a spin $l$ particle exchange, the coefficient $g_l^2$ must be positive in order for the theory to be unitary. This requirement will give constraints on what $M(s,t)$ can be. Also, in the case of chiral string amplitudes, this requirement will give bounds on the chiral parameter $\lambda$. For more information on constraining amplitudes using this method, see [10, 11].

The 4-dilaton amplitude for the chiral string is

$$M_{\text{4-dilaton}} = \lambda^2(st + tu + us)^2 C(\lambda; s, t, u).$$

As usual, this amplitude has poles at $s = 4n, \frac{4n}{\lambda}$. We shall expand $\lambda$ in the field theory limit, namely $\lambda = -1 + \epsilon$.

The expansions are

$$\operatorname*{Res}_{s=4n} M_{\text{4-dilaton}} = c_1(x)\epsilon + c_2(x)\epsilon^2 + \cdots \quad \text{and} \quad \operatorname*{Res}_{s=\frac{4n}{\lambda}} M_{\text{4-dilaton}} = d_1(x)\epsilon + d_2(x)\epsilon^2 + \cdots$$

We have

$$c_1(x) = \frac{64(-1)^n(x^2+3)^2}{(n-1)!(x^2-1)}, \quad d_1(x) = \frac{32(-1)^n(x^2+3)}{n!(x^2-1)}$$

The theory is unitary (to the order of $\epsilon$) if the expansions of $c_1, d_1$ in terms of Gegenbauer polynomial all have negative coefficients.

Unfortunately, the present situation is not the case. The expansion of $c_1(x), d_1(x)$ do not always contain pure negative coefficients. A more obvious problem for $c_1, d_1$ is that they are not polynomials of $x$, meaning that their expansion in terms of Gegenbauer polynomials is infinite. This problem implies that there are exchanges with arbitrary large spins for any fixed mass. The existence of these states contradicts the classification of the conformally invariant states we found in the previous sections. The same problem persists for general $\lambda$ except for $\lambda = \pm 1$. This implies that these amplitudes are pathological. This implication is perhaps not surprising except for $(\lambda = 0, \infty)$, since that for all other $\lambda$, non-conformal states are exchanged during the scattering. These states can easily break the unitarity of the theory.

Although the amplitude we derived above is not physical, this amplitude can still serve as an important tool for studying the relationships between ordinary string amplitudes and particle amplitudes, since we generalized our amplitude with a parameter $\lambda$ that incorporates both the particle theory (the low energy effective field theory) and the string theory (the high energy theory) as specific cases. The parameter $\lambda$ essentially acted as a parameter that interpolates between the low energy and high energy theory. This interpolation occurs in a very natural way since for arbitrary $\lambda$, the chiral string theory for that $\lambda$ correspond to an actual theory, and its properties can be studied in further detail.

# 6 Conclusions

In this paper, we formulated a new class of string amplitudes (and their corresponding spectra) by changing the boundary conditions of string propagators. The different boundary conditions are parametrized by a complex number $\lambda$ and are inspired by duality transformation of 4D Maxwell equations. The amplitudes and spectrum are of normal string theory when $\lambda = 1$. When $\lambda = -1$, the spectrum truncates to finite number and the amplitudes becomes a field theory (supergravity) amplitude.

For general $\lambda$, the chiral string theory can be interpreted as having different tension for left-moving and right-moving modes. In particular, the ratio between the left-moving tension and right-moving tension $\alpha_R/\alpha_L$ is $\lambda$. One can determine whether a particular state for a particular $\lambda$ is conformally invariant by calculating the OPE with respect to the energy-momentum tensor.

We classified different conformally invariant states for any $\lambda$ and we calculated the 4-point amplitude as a function of $\lambda$. The poles of the amplitudes correspond to the states are exchanged during the scattering. One can compare the states exchanged with the list of conformally invariant states and finds that except for $\lambda = 0, \infty, \pm 1$, all other states will exchange non-conformal states, indicating the amplitudes are likely not unitary except for these cases.

One can go a step further and check the unitarity by expanding the residue (of a particular pole) of the amplitude in terms of Gegenbauer polynomials. The necessary and sufficient conditions for unitarity is that the coefficients are all negative. We found that this condition is not true except for $\lambda = \pm 1$, which correspond to the ordinary string theory and the (truncated) field theory.

Although all possible $\lambda$ except for the ones we have previously known are unphysical, these amplitudes are still interesting as they can serve as a tool for studying the relationship between field theory amplitudes and string theory amplitudes. One interesting aspect is renormalizability. As we have found, in the fixed angle and large energy region, the 4-point tree amplitude are exponential decaying if $\text{Re}\lambda > -1$ and exponentially divergent if $\text{Re}\lambda < -1$. On the threshold $\text{Re}\lambda = -1$, the amplitude is polynomially divergent.

Thus, one should be able to calculate the loop amplitudes for general $\lambda$ and see it diverges in the $\lambda = -1$ limit. Since we have the toolset to study many aspects of the theory for arbitrary $\lambda$ (the propagator, spectrum, effective action, etc.), our finding might shed some light on the physical origin of these divergences of (super)gravity.

In general, our formalism provides a new tool to calculate supergravity amplitudes. Just as the CHY method introduced in [1], our method gives a stringy derivation for field theory amplitudes, and it is even more powerful since it is not limited to the tree amplitudes.

Finally, it should be noted that the propagator given in this paper does not exhaust all the possible boundary conditions. There might be new boundary conditions (or even infinite of them) that will result in a unitary and conformal amplitude.

# Acknowledgments

WS thanks Nathan Berkovits for discussion, and WS is supported by NSF grant PHY-1915093. ML would like to thank FAPESP grant 2019/17805-8 for financial support. YPW would like to thank Renann Lipinski Jusinskas for discussion.

# A  A classification of conformally invariant states for chiral strings

In summary, the spectrum of bosonic chiral strings in different $\lambda$ can be split into nine cases:

| case | $\lambda$ | states | mass ($M^2$) |
|---|---|---|---|
| I | irrational | $(1,1)$ | $0$ |
| II | $1$ | $(m,m)$ | $4(m-1)$ |
| III | $p/q > 0$ exclude II | $(kq+1, kp+1)$ | $4kq$ |
| IV | $-1$ | $(2,0), (1,1), (0,2)$ | $4, 0, -4$ |
| V | $-1/n$ | $(1,1), (0, n+1)$ | $0, 4n$ |
| VI | $-n$ | $(1,1), (n+1, 0)$ | $0, -4$ |
| VII | $p/q < 0$ exclude IV - VI. | $(1,1)$ | $0$ |
| VIII | $0$ | $(m, 0)$ | $4(m-1)$ |
| IX | $\lambda \to \infty$, $\lambda \alpha' = 1$ | $(0, m)$ | $4(m-1)$ |

In this table, $k, m = 0, 1 \cdots$ while $n = 2, 3 \cdots$. The nine cases are split in a way that they either have 1, 2, 3 or infinite states and they either have Tachyon or not. II is just the ordinary string theory, while IV is the chiral strings discussed in [6] and V is the case discussed in Jusinskas paper [12].

# B  The kinematic factor of string theory amplitude

The 4-point massless amplitudes for both bosonic and supersymmetric (only the NS-NS sector) chiral string theories can be expressed in a unified way. For more information on the derivation of the kinematic factors, see [8, 9].

$$M_4(1,2,3,4) = \frac{\kappa^2}{128} e^1_{\alpha\mu} e^2_{\beta\nu} e^3_{\gamma\rho} e^4_{\delta\sigma} K_R^{\alpha\beta\gamma\delta} K_L^{\mu\nu\rho\sigma} C(\lambda; s, t, u).$$

$\kappa = \sqrt{32\pi G_{10}}$ is the 10-dimensional gravitational coupling constant. As mentioned in section 5, $e_{\mu\nu}$ is the combined polarization tensor of graviton, 2-form tensor and the dilaton. The kinematic factor $K_{R,L}^{\alpha\beta\gamma\delta}$ can be constructed from the following building blocks:

$$D_{i\;\mu\nu}^{\alpha} \equiv \delta^{\alpha}_{[\mu|} k_{i\;|\nu]}, \quad i = 1, 2, 3, 4.$$

Bosonic amplitude $K_L$ can be written as

$$K_{L\;\alpha\beta\gamma\delta} = K_{0\;\alpha\beta\gamma\delta} + \alpha' K_{1\;\alpha\beta\gamma\delta} + \alpha'^2 stu K_{2\;\alpha\beta\gamma\delta},$$

where

$$K_0^{\alpha\beta\gamma\delta} = 4D_{1\ \mu\nu}^{\alpha}D_{2\ \nu\sigma}^{\beta}D_{3\ \sigma\rho}^{\gamma}D_{4\ \rho\mu}^{\delta} - D_{1\ \mu\nu}^{\alpha}D_{2\ \nu\mu}^{\beta}D_{3\ \sigma\rho}^{\gamma}D_{4\ \rho\sigma}^{\delta}$$
$$+ \ 2 \text{ permutations}.$$

The two additional permutations replace the order 1234 with 1342 and 1243.

$$K_1^{\alpha\beta\gamma\delta} = 4(D_{1\ \mu\nu}^{\alpha}D_{4\ \mu\nu}^{\delta})(k_1 - k_4)_\tau D_{2\ \tau\sigma}^{\beta}D_{3\ \sigma\lambda}^{\gamma}k_{4\lambda} + 8D_{1\ \nu[\mu}^{\alpha}D_{2\ \sigma]\nu}^{\beta}D_{3\ \sigma\rho}^{\gamma}k_4^{\rho}D_{4\ \mu\tau}^{\delta}k_1^{\tau}$$
$$+ \ 3 \text{ permutations}.$$

The following additional three permutations replace 1234 by 2341, 3412 and 4123.

Finally,

$$K_2^{\alpha\beta\gamma\delta} = -2\left(\frac{D_{1\ \mu\nu}^{\alpha}D_{2\ \nu\mu}^{\beta}D_{3\ \sigma\rho}^{\gamma}D_{4\ \rho\sigma}^{\delta}}{s(1+\alpha's)} + \frac{D_{1\ \mu\nu}^{\alpha}D_{2\ \nu\mu}^{\beta}D_{3\ \sigma\rho}^{\gamma}D_{4\ \rho\sigma}^{\delta}}{t(1+\alpha't)} + \frac{D_{1\ \mu\nu}^{\alpha}D_{2\ \nu\mu}^{\beta}D_{3\ \sigma\rho}^{\gamma}D_{4\ \rho\sigma}^{\delta}}{u(1+\alpha'u)}\right)$$

$K_R$ has the exact same expression as $K_L$ except that $\alpha'$ is replaced by $\lambda\alpha'$. For the superstring theory, only $K_0$ are present.

Observe that

$$e_{\alpha\beta}D_{i\ \mu\nu}^{\alpha}D_{i}^{\beta\rho\sigma} = k_{i\ [\mu}k_i^{[\rho}e_{\nu]}^{\sigma]} \equiv W_{[\mu\nu]}^{\ [\rho\sigma]},$$

where $W$ is the Weyl tensor, which is manifestly gauge invariant. Using this observation, the kinematic factor $K_R K_L$ can be written as a product of the 4 Weyl tensors or its derivatives. Each of the 4 Weyl tensors is constructed from the momentum of external states.